\documentclass[a4paper,12pt]{article}
\title{\LARGE\bf $SLE_\kappa$ growth processes\\ and conformal field theories}
\date{}
\author{}

\usepackage[final]{graphicx}
\begin{document}
\maketitle

\vspace{-1.2cm}

\centerline{\large Michel Bauer\footnote[1]{Email:
    bauer@spht.saclay.cea.fr} and Denis Bernard\footnote[2]{Member of
    the CNRS; email: dbernard@spht.saclay.cea.fr}} 

\vspace{.3cm}

\centerline{\large Service de Physique Th\'eorique de Saclay}
\centerline{CEA/DSM/SPhT, Unit\'e de recherche associ\'ee au CNRS}
\centerline{CEA-Saclay, 91191 Gif-sur-Yvette, France}


\vspace{1.0 cm}

\begin{abstract}
  $SLE_\kappa$ stochastic processes describe growth of random curves
  which, in some cases, may be identified with boundaries of two
  dimensional critical percolating clusters. By generalizing
  $SLE_\kappa$ growths to formal Markov processes on the central
  extension of the 2d conformal group, we establish a connection
  between conformal field theories with central charges
  $c_\kappa=\frac{1}{2}(3\kappa-8)(\frac{6}{\kappa}-1)$ and zero modes
  -- observables which are conserved in mean -- of the $SLE_\kappa$
  stochastic processes.

\end{abstract}

\vskip 1.5 truecm


Critical phenomena are characterized by large scale fluctuations.
Conformal field theories \cite{bpz} are powerful tools for analyzing
their multifractal properties, leading for instance to exact results
concerning scaling behavior of geometrical models, see eg.
refs.\cite{nienhuis}. An alternative probabilistic approach has
recently been introduced \cite{schramm0}. It consists in formulating
conformally covariant processes, so-called $SLE_\kappa$ processes,
based on Loewner's equation.  Among the many conformal field theory results,
including crossing percolation probabilities \cite{cardy} or
multifractal distributions of electrostatic potential near conformally
invariant fractal boundaries \cite{duplan}, some have been rederived
in the $SLE_\kappa$ framework, see eg refs.\cite{schramm,werner} for a
review. However, the precise connection between $SLE_\kappa$ models
and conformal field theories in the sense of \cite{bpz} remains
elusive, although $SLE_6$ has been identified with 2d percolation,
with central charge $c=0$, $SLE_2$ with loop erased random walks, with
$c=-2$, and $SLE_{8/3}$ with self-avoiding random walks, with $c=0$
also. But see ref.\cite{bertrand}.

The aim of this letter is to establish a direct relation between
$c\leq 1$ conformal field theories and $SLE_\kappa$ models by
extending them to processes in the Virasoro group.  This link goes
through the construction (using conformal field theory) of observables
which are conserved in mean during the $SLE_\kappa$ growth.  It bears
some analogy with turbulent passive advections \cite{zero,slow}.

\vskip 1.0 truecm 

{\bf $SLE_\kappa$ processes.}
Stochastic Loewner evolution $SLE_\kappa$ are growth processes \cite{schramm0}
defined via conformal maps which are solutions of Loewner's equation:
\begin{equation}
\partial_t g_t(z)=\frac{2}{g_t(z)-\xi_t}\ ,\quad g_{t=0}(z)=z
\label{loew}
\end{equation}
When $\xi_t$ is a smooth real-valued function, the map $g_t(z)$ is the
uniformizing map for a simply connected domain $D_t$ of the upper half
plane ${\bf H}$, ${\rm Im}z>0$. The map $g_t(z)$, normalized by
$g_t(z)=z +2t/z + \cdots$ at infinity, is well-defined up to
the time $\tau_z$ for which $g_{\tau_z}(z)=\xi_{\tau_z}$.  Notice that
${\rm Im}g_t(z)$ is a decreasing function of time on ${\bf H}$.
Following refs.\cite{schramm0,schramm,werner}, define $K_t=\{z\in{\bf
  H}:\ \tau_z\leq t\}$.  They form an increasing sequence of sets,
$K_{t'}\subset K_t$ for $t'<t$, and for smooth enough driving source
$\xi_t$, they are simple curves staggering along ${\bf H}$.  The
domain $D_t$ is ${\bf H}\setminus K_t$.

$SLE_\kappa$ processes are defined \cite{schramm0} by choosing a Brownian motion as
driving parameter in the Loewner's equation: 
$\xi_t=\sqrt{\kappa}\, B_t$ with $B_t$ a two-sided normalized Brownian
motion and $\kappa$ a parameter. The growth processes are then that of the
sets $K_t$.

\vskip 1.0 truecm 

{\bf Lifted $SLE_\kappa$ processes.}  
Let $L_n$ be the generators of the Virasoro algebra $vir$ --
the central extension of the Lie algebra of conformal transformations --
with commutation relations
$[L_n,L_m]=(n-m)L_{n+m}+\frac{c}{12}n(n^2-1)\delta_{n+m,0}$.
Let ${\cal V}ir$ be the formal
group obtained by exponentiating elements of $vir$.  We
define a (formal) stochastic Markov process on ${\cal V}ir$ by the first order
stochastic differential equation generalizing random walks on Lie
groups:
\begin{equation}
G_t^{-1}\, dG_t = -2dt\,L_{-2} + d\xi_t\, L_{-1}
\label{slevir}
\end{equation}
where $d\xi_t\equiv w_tdt$, so that $w_t$ is Gaussian with white-noice
covariance $\langle w_t\, w_s\rangle = \kappa \delta(t-s)$.  With
initial condition $G_{t=0}=1$, the elements $G_t$ belong to the formal
group ${\cal V}ir_-$ obtained by exponentiating the generators $L_n$,
$n<0$, of negative grades of the Virasoro algebra.  We may order
factors in ${\cal V}ir_-$ according to their grades $n$, so that group
elements $G$ are presented in the form $\cdots e^{x_2 L_{-2}}e^{x_1
  L_{-1}}$. Eq.(\ref{slevir}) turns into a family of ordinary
stochastic differential equations for the $x_k$'s, with a
probabilistic measure induced by that of the Brownian motion.

The connection with the previously introduced $SLE_\kappa$ growth
process emerges from the action induced by the above flows on
conformal fields. Recall from ref.\cite{bpz}, that conformal primary
fields $\phi_\Delta(z)$ are $z$-dependent operators acting on
appropriate representations of the Virasoro algebra and transforming as
forms of weight $\Delta$ under conformal transformations. They satisfy
the following intertwining relations:
$[L_n,\phi_\Delta(z)]=\ell^\Delta_n\cdot \phi_\Delta(z)$ with
$\ell_n^\Delta \equiv z^{n+1}\partial_z + (n+1)\Delta z^n$.  As a
consequence, the group ${\cal V}ir$ acts on primary fields by
conformal transformations and, in particular for the flows
(\ref{slevir}), one has:
$$
G^{-1}_t\, \phi_\Delta(z)\, G_t = [\partial_z \hat g_{t}(z)]^\Delta\,
\phi_\Delta(\hat g_{t}(z))
$$
where $\hat g_t(z)\equiv g_t(z)-\xi_{t}$ with $g_t(z)$ solution of the
Loewner's equation.  This means that the conformal transformations
induced by the lifted $SLE_\kappa$ flows on the primary fields reduce
to that of the conformal maps of the $SLE_\kappa$ processes.

Alternatively, the $SLE_\kappa$ stochastic equation may be represented as
$\hat g_t(z)=H_t\,z\,H_t^{-1}$ with
$H^{-1}_t dH_t = 2dt\,\ell_{-2}^0-d\xi_t\,\ell_{-1}^0$.
Indeed, equation (\ref{loew}) translates into $\partial_t \hat
g_t(z)=2/ \hat g_t(z) - w_t$ for $\hat g_t(z)=g_t(z)-\xi_t$ which is
then equivalent to the commutation relation $[H^{-1}_t\partial_t
H_t,z]=2/z-w_t$ among differential operators.  The group ${\cal
  V}ir_-$ may be presented as the group of germs of meromorphic
functions with a pole at infinity such that
$f(z)=z(1+a_1/z+a_2/z^2+\cdots)$ with $a_k$ as coordinates, similar to
Fock space coordinates. The Virasoro generators are then differential
operators in the $a_k$'s.

\vskip 1.0 truecm

{\bf Time evolution and Fokker-Planck equations.}  
Any observable of the random process $G_t$ may be thought of as
function on ${\cal V}ir$, and we use formal rules extending those
valid in finite dimensional Lie groups.  We denote by ${\bf E}[
F(G_t)]$ the expectation value of the observable $F(G_t)$.

Our main result is the following representation
of their time evolution:
\begin{equation}
\partial_t\, {\bf E}[
F(G_t) ] = {\bf E}[ -2 \nabla_{-2}F(G_t) + \frac{\kappa}{2} 
\nabla_{-1}^2 F(G_t) ]
\label{slevol}
\end{equation}
where $\nabla_n$ are the left invariant vector fields associated to the
elements $L_n$ in $vir$ defined by 
$(\nabla_n F)(G) = \frac{d}{du} F(G\, e^{uL_n})\vert_{u=0}$
for any appropriate function $F$ on ${\cal V}ir$.

Hints for the proof of eq.(\ref{slevol}) are as follows.  By
definition of the lifted $SLE_\kappa$ flows, the mean value of any
test function of such flows evolves according to $\partial_t {\bf E}[
F(G_t) ] = {\bf E}[-2 \nabla_{-2}F(G_t) + w_t
\nabla_{-1}F(G_t)]$. Since $w_t$ is a Gaussian variable with white-noice
covariance, the last term may be identified with $\kappa {\bf E}[
\nabla_{-1} \delta F(G_t)/\delta w_t]$. So one has to compute
$\delta F(G_t)/\delta w_t$. Small variation of eq.(\ref{slevir})
implies that 
$\frac{d}{dt}(\frac{\delta G_t}{\delta w_s}\, G^{-1}_t)=
(G_s L_{-1} G^{-1}_s) \delta(t-s).$
As a consequence, we get 
$\delta F(G_t)/\delta w_t=\frac{1}{2} \nabla_{-1} F(G_t)$ and
eq.(\ref{slevol}).

As for any Markov process, the time evolution of the probability
distribution functions (Pdf) of the lifted $SLE_\kappa$ processes are governed
by Fokker-Planck equations whose hamiltonians are the generators of the
semi-groups specifying the processes. These Pdf's, denoted ${\cal P}_t(G)$,
may be written as the averages of the point Dirac measures localized
on $G$: ${\cal P}_t(G)\equiv {\bf E}[ \delta_G(G_t)]$.  
Their time evolution is:
\begin{equation}
\partial_t {\cal P}_t(G) ={\cal H}\cdot{\cal P}_t(G),\quad 
{\cal H}\equiv 2 \nabla_{-2} + \frac{\kappa}{2}\nabla_{-1}^2
\label{slefokker}
\end{equation}
This follows from eq.(\ref{slevol}) with $F(G_t)=\delta_G(G_t)$, using the fact
that the Lie derivatives of $\delta_G(G_t)$ with respect to $G_t$ are 
the opposite of its Lie derivatives with respect to $G$.

By eq.(\ref{slefokker}), the Pdf's are time transported by the
Fokker-Planck hamiltonian ${\cal H}$ so that we expect
${\cal P}_t(G) = \exp t{\cal H}\cdot {\cal P}_{t=0}(G)$.
Alternatively, the probability transitions from $G_0$ at time $t_0$
to $G$ at time $t>t_0$ are the kernels of the operator $\Big( \exp
(t-t_0){\cal H} \Big)_{G,G_0}$.  A similar derivation remains valid if
we consider the stochastic flows $x_t=G^{-1}_t\cdot x_{0}$ induced by
eq.(\ref{slevir}) on any representation of ${\cal V}ir$ .  Of course,
these stochastic processes are calling for a more mathematical precise
description, along the lines of refs.\cite{schramm0, schramm,werner}.

\vskip 1.0 truecm

{\bf Zero modes and null vectors.} 
Since left invariant Lie derivatives form a representation of $vir$,
equation (\ref{slevol}) may be written as:
\begin{equation}
\partial_t\, {\bf E}[F(G_t)] = {\bf E}[ {\cal H}^T\cdot F(G_t) ],\quad
{\cal H}^T \equiv -2 L_{-2} +\frac{\kappa}{2}\, L_{-1}^2
\label{slehamil}
\end{equation}
with the Virasoro generators $L_n$ acting on the appropriate
representation space.  By definition a
zero modes $F_\omega$  is an eigenvector of ${\cal H}^T$ with zero
eigenvalue: ${\cal H}^T\cdot F_\omega=0$. As a consequence 
a zero mode is an observable conserved in mean.

To construct such conserved quantities we look for such $F_\omega$'s
among zero modes annihilated by the $L_n$, $n>0$, ie. among the
so-called highest weight vectors, $L_n F_\omega=0$, $n>0$, with
given conformal dimension $h$, $L_0 F_\omega=h F_\omega$.
We now demand under which conditions ${\cal H}^T F_\omega$ is
again a highest weight vector.  We have
$$ [L_n, {\cal H}^T] = (-2(n+2)+\frac{\kappa}{2}n(n+1))L_{n-2}
+ \kappa (n+1)L_{-1}L_{n-1} - c\delta_{n,2}
$$
Hence, $L_n {\cal H}^T F_\omega=0$ for any $n\geq 3$, 
but demanding $L_1 {\cal H}^T F_\omega=0$ requires
$2\kappa h=6-\kappa$, whereas $L_2 {\cal H}^T F_\omega=0$
imposes $c=h(3\kappa-8)$.

Thus, for these values of $c$ and $h$, ${\cal H}^T F_\omega$ is a
highest weight vector, a so-called null vector, which can be
consistently set to zero, see eg.\cite{bpz}. In other words,
$F_\omega$ is such that ${\cal H}^T F_\omega\simeq 0$ if and only
if the Virasoro central charge is adjusted to
\begin{equation}
c_\kappa = \frac{1}{2}\Big(3\kappa-8\Big)\Big(\frac{6}{\kappa}-1\Big)
= 1 - 6 \Big(\frac{2}{\sqrt{\kappa}}-\frac{\sqrt{\kappa}}{2}\Big)^2
\label{dingdong}
\end{equation}
and the conformal weight to $h_\kappa=\frac{1}{2}(\frac{6}{\kappa}-1)$.
Furtheremore, the stationary property of the mean of 
$F_\omega$ is a consequence of the existence of a null 
vector in the corresponding Virasoro Verma module.  The analogy with
conserved modes in turbulent passive advection \cite{zero,slow} is
particularly striking.

Such zero modes may be constructed by considering the orbit of the 
highest weight vector $\omega$ of conformal dimension $h_\kappa$ in
the $c_\kappa$ conformal field theory and defining
$F_\omega(\omega)= G_t\cdot \omega$ so that
$$
\partial_t\, {\bf E}\,[ G_t \cdot \omega ] =0
$$
All components of the vector ${\bf E}\,[ G_t \cdot \omega ]$ are
conserved. Similarly, since the conformal field $\phi_x(z)$
associated to any state $x$ in the Virasoro module with highest weight
vector $\omega$ depends linearly on $x$, the matrix elements of $ {\bf
  E}\,[ \phi_{G_t\cdot \omega} (z) ]$ are also conserved.  \bigskip

In conclusion, we have established a direct link between observables
conserved in mean and  conformal null vectors.
This opens a route to the study of probabilistic
properties of growth processes using conformal field theory.
The above central charge coincides with that of the
conformal field theory expected to be associated with the
$SLE_\kappa$ process. It is invariant under the duality
transformation $\kappa \to 16/\kappa$, it vanishes for $\kappa=6$ as
expected for percolation, and the self-dual point $\kappa=4$
corresponds to the $c_\kappa=1$ free field theory.  

More on the algebraic and geometrical aspects as well as on
applications of this construction will be described elsewhere
\cite{avenir}.

\vskip 1.0 truecm

{\bf Acknowledgement:} We thank Antti Kupiainen for explanations 
concerning the $SLE$ processes.


\end{document}